\documentclass[sigconf]{acmart}
\usepackage{amsmath,amssymb,amsfonts}
\usepackage{graphicx}
\usepackage{textcomp}
\usepackage{lipsum}
\usepackage[absolute]{textpos}
\usepackage{mathtools}
\usepackage[T1]{fontenc}
\usepackage{enumitem,kantlipsum}

\title{Demo Abstract: {\textsc{VoltKey}}: Using Power Line Noise for Zero--Involvement Pairing and Authentication }

\author{ Jack West, Tien VoNguyen, Isaac Ahlgren, Iryna Motyashok, George K. Thiruvathukal, Neil Klingensmith} 
    \affiliation{ 
      \institution{University of Loyola Chicago}
      \streetaddress{1052 West Loyola Avenue}
      \city{Chicago}
      \state{Illinois}
      \country{USA}
      }
    \email{
      {jwest1@, tvonguyen@, iahlgren@, imotyashok@, gkt@cs.,neil@cs.}luc.edu
      }
 
\author{Kyuin Lee, Dong He, Younghyun Kim, Suman Banerjee}
\affiliation{%
  \institution{University of Wisconsin--Madison}
  \streetaddress{1415 Engineering Dr}
  \city{Madison}
  \state{Wisconsin}
  \country{USA}}
\email{kyuin.lee@wisc.edu, dhe28@wisc.edu}

\keywords{Privacy; Mobile Systems; IoT; Hypervisors; Real-time}

\begin{document}

\copyrightyear{2019} 
\acmYear{2019} 
\setcopyright{none}
\acmConference[IPSN '19]{The 20th International Workshop on Mobile Computing Systems and Applications}{February 27--28, 2019}{Santa Cruz, CA, USA}

\begin{abstract}
We present \textsc{VoltKey}, a method that transparently generates secret keys for colocated devices, leveraging spatiotemporally unique noise contexts observed in commercial power line infrastructure.
\textsc{VoltKey} extracts randomness from power line noise and securely converts it into an authentication token.
Nearby devices which observe the same noise patterns on the powerline generate identical keys.
The unique noise pattern observed only by trusted devices connected to a local power line prevents malicious devices without physical access from obtaining unauthorized access to the network.
\textsc{VoltKey} is implemented inside of a standard USB power supply as a platform-agnostic bolt-on addition to any IoT or mobile device or any wireless access point that is connected to the power outlet.
\end{abstract}

\maketitle

\section{Introduction}

As breaches in apps, websites, and credit card processing networks have become commonplace, it is increasingly clear that the best kind of password is one that doesn't exist.
User-managed passwords are perhaps the most important and widespread security weakness of existing apps.
Reliance on passwords is responsible for large-scale identity theft, fraud, and even election manipulation.
Usability is an important consideration for personal IoT systems, mobile apps, and web accounts that are deployed and maintained by non-professional users.
In particular, one of the paramount concerns that have continued to vex researchers is the question of how to quickly, securely, and effortlessly establish a common security key between a newly introduced device and an existing network and to subsequently manage the established connection securely.

\begin{figure}
\centering
  \includegraphics[width=0.45\textwidth]{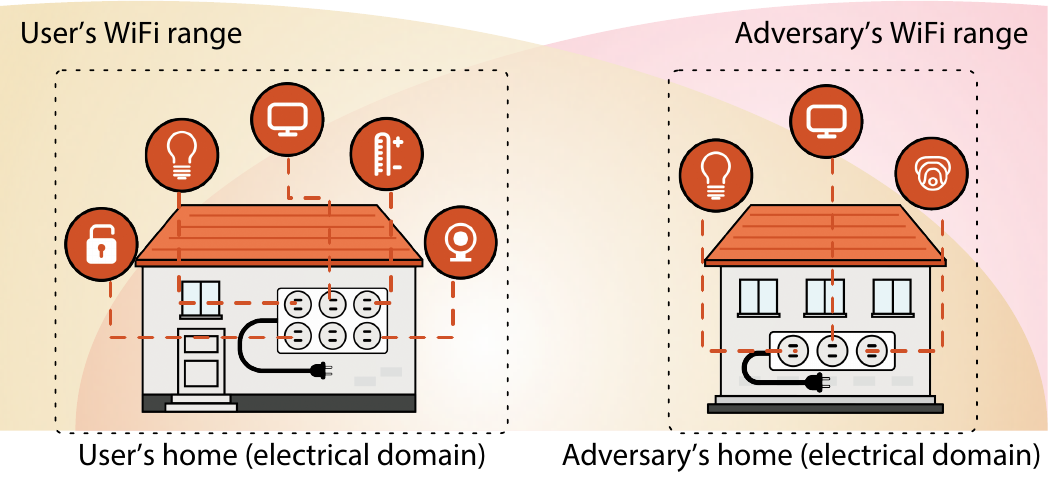}
  \vskip -10pt
  \caption{System and threat models of \textsc{VoltKey}. A number of IoT devices are installed in each home. As long as anyone is in range of a router they can potentially access it.}~\label{fig:introduction}
 \vskip -10pt
\end{figure}

\emph{Context-based pairing and authentication} is a promising solution to this challenge.
It exploits spatiotemporal randomness in the ambient environment often called \emph{contextual information}~\cite{Fomichev-CST18}.
The presence of common contextual information is evidence that the devices are located in the same place at the same time, which implies that they legitimately belong to the same user.
The keys generated from contextual information can, therefore, be used to establish initial trust (as a pairing key) and to protect subsequent communication (as a cryptographic key).
This eliminates the need for human involvement for making, entering, and managing a secret key, which dramatically improves the overall usability of IoT systems.
In addition, the time-varying nature of contextual information also allows devices to use a new key for each pairing attempt or periodically update the cryptographic key, which significantly reduces the attack window for adversarial agents.


In this work, we demonstrate a key generation method named \textsc{VoltKey}~\cite{voltkey}, which harvests correlated random noise for wall power outlets to generate authentication keys.
More specifically, \textsc{VoltKey} takes advantage of the fact that devices that are powered by colocated electrical outlets, or those that are within the same \emph{authenticated electrical domain}, observe similar  \emph{noise fingerprints} caused by the nearby electrical environment which is temporally and spatially unique.
\textsc{VoltKey} is embedded in standard USB power supplies that are pervasively used in personal and domestic IoT and mobile devices.
For the reason that, it exploits standard power line infrastructure that is ubiquitously available virtually everywhere, \textsc{VoltKey} does not require additional supporting infrastructure for installation.
Using \textsc{VoltKey}, devices that wish to associate with one another can simply be plugged into an existing power outlet to automatically generate (and periodically regenerate) a unique key and associate themselves with no involvement from the user.

In \textsc{VoltKey}-enabled networks, a device's ability to authenticate itself is dependent on its physical proximity to the host access point.
The boundaries of an authenticated electrical domain are determined by the electrical interconnect of the site.

\section{\textsc{VoltKey} Hardware Design}
\begin{figure}
\centering
  \includegraphics[width=\hsize]{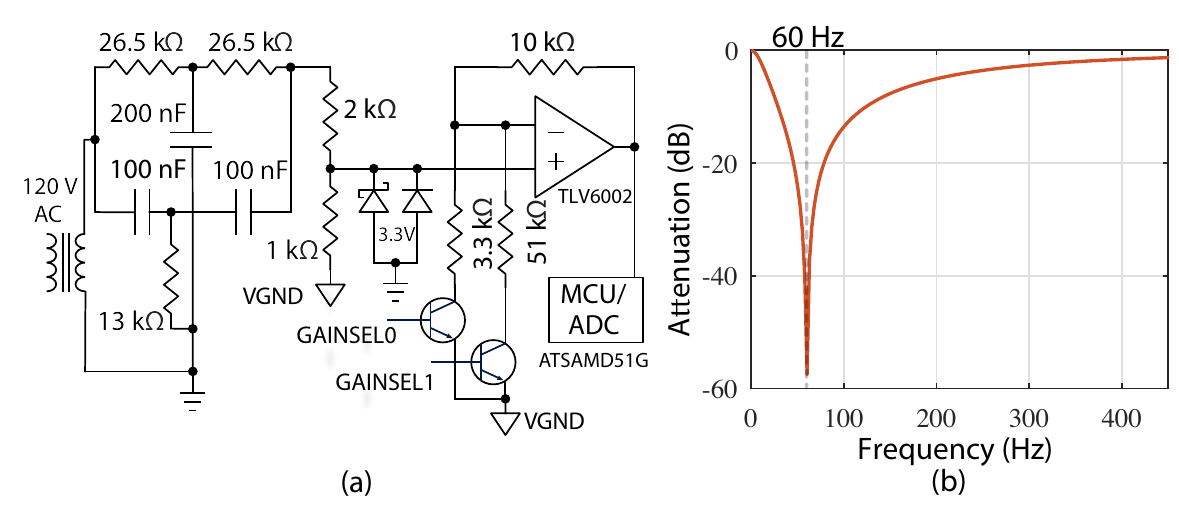}
  \vskip -10pt
  \caption{(a) \textsc{VoltKey}'s analog front-end schematic. (b) Freq. response of the twin-T notch filter used in our prototype.}
  \label{fig:hardware}
  \vskip -10pt
\end{figure}
We design \textsc{VoltKey} as a modular addition to standard USB shipped with IoT devices.
In addition to supplying power, the module also generates keys from superimposed noise on the power line and transmits the keys to the device over a wired interface for pairing and authentication purposes.
\textsc{VoltKey} consists of two main components: (1) the analog input circuitry for filtering and amplifying power line noise and (2) a microcontroller unit (MCU) which includes an analog-to-digital converter (ADC) for noise measurement and key extraction procedures. 

\paragraph{Analog front-end}
The analog front-end, illustrated in Fig.~\ref{fig:hardware}(a), consists of an isolation transformer, a twin-T notch filter, and a differential amplifier.
The purpose of this circuit is to amplify high-frequency noise from the power line and attenuate the 60~Hz fundamental.
The transformer steps the 120~V AC power signal (between hot and neutral) down to a lower voltage and isolates the \textsc{VoltKey} circuitry from the power line.
Our prototype uses a split-core transformer with two secondary coils: one to generate power for our circuitry and the host and another to measure noise.
We do not want to measure noise on the same transformer tap that we use to generate power because the noise from \textsc{VoltKey} 's digital components may corrupt the power line noise measurement.
As illustrated in Fig.~\ref{fig:hardware}(b), the twin-T notch filter attenuates the 60~Hz fundamental frequency component from the voltage waveform.
The 60~Hz component is an unwanted signal in the context of \textsc{VoltKey} because its harmonics carry a deterministic signal that repeats almost identically from period to period.
Therefore, attenuating it improves the signal-to-noise ratio (SNR).
After the signal has been filtered, it has an amplitude of 200--300~mV and an average value of 0~V.
The amplifier's job is to shift and amplify the filtered signal so that its range is within 0--3.3~V, the limits of the ADC.
The diodes at the end of the filter clip are filtered analog voltage waveforms between 0--3.3~V to avoid damaging the op-amp and the ADC of the MCU.
We use an op-amp to generate a virtual ground of 0.7~V, and the output of the twin-T notch filter is referenced to the virtual ground using a voltage divider (immediately to the left of the diodes in Fig. \ref{fig:hardware}(a)).
The amplitude of the noise varies considerably depending on active electrical loads.
When the signal amplitude is too small compared to the ADC dynamic range, we may get a poor measurement due to a large quantization error; if it is too big, the peaks of the noise will be clipped by the diodes and lost.
To deal with this issue, we build a adjustable gain amplifier to allow software to dynamically adapt to changing noise conditions, adjusting the gain accordingly.
The adjustable gain amplifier is built from a standard configuration of a non-inverting op-amp circuit with bipolar junction transistors in the feedback loop between the inverting input and VGND.
The \texttt{GAINSELx} signals are connected to the microcontroller's GPIO lines via bias resistors, allowing software to modify the amplifier's gain.

\paragraph{MCU and ADC}
\textsc{VoltKey} uses a low-cost MCU to measure and process the voltage signal on the power outlet.
Our hardware prototype is equipped with the Microchip's ATSAMD51, a 32-bit ARM-Cortex M4~\cite{samd51}, running at 120~MHz with an on-chip ADC capable of sampling rate of up to 1~MSPS at a 12-bit resolution.
The MCU also has a USB device functionality which can be used to transfer the computed keys to the host over a virtual COM port (serial) interface.
The MCU we chose is considerably more powerful than necessary---\textsc{VoltKey}'s application uses very little memory and can run on a low-power processor.

\paragraph{Key Generation Protocol}

\begin{enumerate}[wide, labelwidth=!, labelindent=0pt]
    \item \textit{Measurement: }Device $B$ (e.g., an IoT device) contacts Device $A$ (e.g., a WiFi access point) to initiate independent power line noise measurement. 
    Note that the sampling clock (time and rate) can vary between $A$ and $B$.
    \item \textit{Sampling rate estimation: }Each device independently estimates the sampling rate of its own ADC sample buffer.
    Let $c_A$ and $c_B$ be the estimated sampling rate of $A$ and $B$.
    \item \textit{Sampling rate matching: }Both devices resample their ADC buffers to make the sampling rates are identical.
    \label{step:sampling_rate}
    \item \textit{Time synchronization: }Device $B$ syncs its measurement time to that of $A$ using a short snippet of samples shared from Device $A$.
    \label{step:time_synchronization}
    \item \textit{Bit sequence extraction: }Both devices independently extract bit sequences from their sample buffers. 
    \item \textit{Reconciliation: }Differences in the extracted bit sequences are corrected by $B$ with publicly exchanged data through key reconciliation stage.  \label{step:reconciliation}
\end{enumerate}

\balance
\bibliographystyle{ACM-Reference-Format}
\bibliography{refs}  


\begin{thebibliography}{3}


\ifx \showCODEN    \undefined \def \showCODEN     #1{\unskip}     \fi
\ifx \showDOI      \undefined \def \showDOI       #1{#1}\fi
\ifx \showISBNx    \undefined \def \showISBNx     #1{\unskip}     \fi
\ifx \showISBNxiii \undefined \def \showISBNxiii  #1{\unskip}     \fi
\ifx \showISSN     \undefined \def \showISSN      #1{\unskip}     \fi
\ifx \showLCCN     \undefined \def \showLCCN      #1{\unskip}     \fi
\ifx \shownote     \undefined \def \shownote      #1{#1}          \fi
\ifx \showarticletitle \undefined \def \showarticletitle #1{#1}   \fi
\ifx \showURL      \undefined \def \showURL       {\relax}        \fi
\providecommand\bibfield[2]{#2}
\providecommand\bibinfo[2]{#2}
\providecommand\natexlab[1]{#1}
\providecommand\showeprint[2][]{arXiv:#2}

\bibitem[\protect\citeauthoryear{{Fomichev}, {\'Alvarez}, {Steinmetzer},
  {Gardner-Stephen}, and {Hollick}}{{Fomichev} et~al\mbox{.}}{2018}]%
        {Fomichev-CST18}
\bibfield{author}{\bibinfo{person}{Mikhail {Fomichev}}, \bibinfo{person}{Flor
  {\'Alvarez}}, \bibinfo{person}{Daniel {Steinmetzer}}, \bibinfo{person}{Paul
  {Gardner-Stephen}}, {and} \bibinfo{person}{Matthias {Hollick}}.}
  \bibinfo{year}{2018}\natexlab{}.
\newblock \showarticletitle{Survey and Systematization of Secure Device
  Pairing}.
\newblock \bibinfo{journal}{\emph{IEEE Communications Surveys Tutorials}}
  \bibinfo{volume}{20}, \bibinfo{number}{1} (\bibinfo{date}{Firstquarter}
  \bibinfo{year}{2018}), \bibinfo{pages}{517--550}.
\newblock
\showISSN{1553-877X}
\urldef\tempurl%
\url{https://doi.org/10.1109/COMST.2017.2748278}
\showDOI{\tempurl}


\bibitem[\protect\citeauthoryear{Inc.}{Inc.}{2018}]%
        {samd51}
\bibfield{author}{\bibinfo{person}{Microchip~Technology Inc.}}
  \bibinfo{year}{2018}\natexlab{}.
\newblock \bibinfo{title}{SAM D5x/E5x Family Data Sheet}.
\newblock
\newblock
\urldef\tempurl%
\url{http://ww1.microchip.com/downloads/en/DeviceDoc/60001507C.pdf}
\showURL{%
\tempurl}


\bibitem[\protect\citeauthoryear{Lee, Klingensmith, Banerjee, and Kim}{Lee
  et~al\mbox{.}}{2019}]%
        {voltkey}
\bibfield{author}{\bibinfo{person}{Kyuin Lee}, \bibinfo{person}{Neil
  Klingensmith}, \bibinfo{person}{Suman Banerjee}, {and}
  \bibinfo{person}{Younghyun Kim}.} \bibinfo{year}{2019}\natexlab{}.
\newblock \showarticletitle{VoltKey: Continuous Secret Key Generation Based on
  Power Line Noise for Zero-Involvement Pairing and Authentication}.
\newblock \bibinfo{journal}{\emph{Proc. ACM Interact. Mob. Wearable Ubiquitous
  Technol.}} \bibinfo{volume}{3}, \bibinfo{number}{3}, Article
  \bibinfo{articleno}{Article 93} (\bibinfo{date}{Sept.} \bibinfo{year}{2019}),
  \bibinfo{numpages}{26}~pages.
\newblock
\urldef\tempurl%
\url{https://doi.org/10.1145/3351251}
\showDOI{\tempurl}


\end{thebibliography}

\end{document}